\documentclass[12pt]{aastex}
\usepackage{emulateapj5}

\usepackage{psfig}
\usepackage{amsmath}

\newcommand{\pr}{\textrm{Pr}}
\newcommand{\data}{\textrm{\,Data\,}}

\newcommand{\pars}{\,\{\theta\}\,}

\shorttitle{ {\it HST} Study of Cl0024+16: II The Cluster Mass
Distribution} \shortauthors{Kneib et~al.}

\begin{document}

\title{A Wide Field Hubble Space Telescope Study of the Cluster
Cl0024+1654 at $z=0.4$ II: The Cluster Mass Distribution}
\author{ Jean-Paul Kneib\altaffilmark{1,2},
Patrick Hudelot\altaffilmark{1},
Richard S. Ellis\altaffilmark{2},
Tommaso Treu\altaffilmark{2},
Graham P. Smith\altaffilmark{2},
Phil Marshall\altaffilmark{3},
Oliver Czoske\altaffilmark{4},
Ian Smail\altaffilmark{5} \&
Priya Natarajan\altaffilmark{6} }

\altaffiltext{1}{Observatoire Midi-Pyr\'en\'ees, UMR5572,
         14 Avenue Edouard Belin,
         31000 Toulouse, France.}
\altaffiltext{2}{Caltech, Astronomy, 105-24, Pasadena, CA 91125, USA.}
\altaffiltext{3}{MRAO, Cavendish Laboratory, Cambridge, CB3 0HA, UK.}
\altaffiltext{4}{Institut f\"ur Astrophysik und Extraterrestrische
Forschung, Auf dem H\"ugel 71, 53121 Bonn, Germany.}
\altaffiltext{5}{Institute for Computational Cosmology, University of Durham, South
Road, Durham DH1 3LE, UK.}
\altaffiltext{6}{Dept. of Astronomy, Yale University, P.O. Box 208101, New Haven,
CT 06250, USA.}

\begin{abstract}
We present a comprehensive lensing analysis of the rich cluster
Cl0024+1654 (z=0.395) based on panoramic sparse-sampled imaging
conducted with the WFPC2 and STIS cameras on board the {\it Hubble
Space Telescope}. By comparing higher fidelity signals in the
limited STIS data with the wider field data available from 
WFPC2, we demonstrate an ability to detect reliably 
weak lensing signals to a cluster radius of $\simeq$5
$h_{65}^{-1}$ Mpc where the mean shear is around 1\%. This enables
us to study the distribution of dark matter with respect to the
cluster light over an unprecedented range of cluster radius and
environments. The projected mass distribution reveals a
secondary concentration representing 30\% of the overall
cluster mass, which is also visible in the distribution of cluster member
galaxies.  We develop a method to derive the projected mass profile of the
main cluster taking into account the influence of the secondary
clump. We normalize the mass profile determined from the shear by
assuming that background galaxies selected with 23$<I<$26 have a
redshift distribution statistically similar to that inferred
photometrically in the Hubble Deep Fields (HDFs). The total mass within
the central region of the cluster is independently determined from strong
lensing constraints according to a detailed model which utilizes
the multiply-imaged arc at $z=1.675$. Combining strong and weak 
constraints, we are able to probe 
the mass profile of the cluster on scales of 0.1 to 5 Mpc thus providing a
valuable test of the universal form proposed by Navarro, Frenk \& White (1997) 
on large scales. A generalized power law fit indicates an asymptotic
3-D density distribution of $\rho\propto\,r^{-n}$ with
$n>$2.4.  An isothermal mass profile is therefore strongly rejected, 
whereas a NFW profile with M$_{200}$= 6.1$^{+1.2}_{1.1}$ 
10$^{14} h_{65}^{-1}$ M$_\sun$ provides a good fit to the lensing data.
We isolate cluster members according to their
optical-near infrared colors; the red cluster light closely traces
the dark matter with a mean mass-to-light ratio of M/L$_K$= 40$\pm 5$
$h_{65}$ M$_\sun$/L$_\sun$. Similar profiles for mass and light on
1-5 Mpc scales are expected if cluster assembly is largely
governed by infalling groups.
\end{abstract}



\section{Introduction}

Clusters of galaxies are the largest dynamically bound systems in
the Universe and act as effective laboratories for studying the
relationship between the distributions of dark and baryonic
matter. In currently popular cold dark matter models, numerical
simulations predict a universal dark matter profile (Navarro,
Frenk, \& White 1997, hereafter NFW), which falls off as
$\rho\propto r^{-3}$ at large radius with a central cusp whose
limiting slope (on scales of 10-100 kpc\footnote{We assume $H_0=$65
km/s/Mpc, $\Omega_m=0.3$, $\Omega_\Lambda=0.7$ throughout.
At the redshift of the cluster $1\arcsec$ corresponds to 5.74 kpc 
(and $1'$ to 0.344 Mpc).  }) is thought to lie between $-1$
(NFW) and $-1.5$ (Moore et al.\ 1998, Ghigna et al.\ 2000).

To date, observational attempts to verify this universal profile in
clusters of galaxies have largely been confined to studies on
small scales. Evidence for a core might support the idea that the
dark matter is warm or self-interacting (Spergel \& Steinhardt
2000). Unfortunately, tests of the NFW profile on 10-100 kpc scales
are complicated by the fact that baryons are dominant in this
regime (Smith et al.\ 2001). Not only must the baryonic component be removed
meticulously via dynamical and photometric data ({\it e.g.} Sand et al.\
2002), but baryonic collapse during cluster formation presumably
also steepens the inner dark matter profile beyond that predicted
in {\it dark matter only} numerical simulations. The interpretation of 
the inner mass profile is also likely be confused by projection effects and
line-of-sight cluster mergers (Czoske et al.\ 2002).

On scales from 100 kpc up to a few Mpc, e.g. in massive clusters
of galaxies, numerical predictions of the dark matter profile
should largely be unaffected by the effects of baryonic collapse.
Moreover, the total mass profile on such scales will largely be
represented by that of the dark component. Thus baryonic contamination 
should not be a major issue. The difficulty in observationally verifying the
form of the NFW profile on large scales lies in securing robust measurements
of the radial profile over a sufficiently wide range in radius.
Substructures may also confuse the derived profiles.

Weak gravitational lensing is the most reliable tool for
quantifying the total mass distribution over a wide range of
cosmic scales (Mellier 1999 and references therein). 
Studies undertaken with the Wide
Field Planetary Camera 2 (WFPC2) on {\it Hubble Space Telescope} ({\it HST})
have determined cluster mass distributions both in the central
parts (Seitz et al.\ 1996) and on larger scales ($<$1.5 Mpc,
Hoekstra et al.\ 1998, 2002). Accordingly, the prospect of measuring the
cluster mass distribution to very large scales via this technique
is promising if the various systematic effects can be understood
and properly evaluated.

Comparing the relative distribution of the baryonic and dark
matter components over a wide range of scales is also of
considerable astrophysical interest. In popular dark matter
models, the stellar component is expected to be biased with
respect to the dark matter depending on its color. Whereas
suitably defined cluster core radii as determined from baryonic tracers
and lensing studies are now thought to be in general agreement
(Allen et al.\ 2001), less is known about the respective
distributions on larger scales (Wilson et al.\ 2001, Gray et al.\
2001, Clowe \& Schneider 2001, 2002). 

This paper is concerned with a comprehensive attempt to detect
weak gravitational lensing signals from the core of a rich cluster
out to $\simeq$5 Mpc (which is roughly three times the virial radius).
As the lensing shear at this radius is expected to be very small
($\approx$1\%), such a study is possible only by using the high
resolution imaging capabilities of {\it HST}.
A major goal is to combine the dark matter
profile derived from weak lensing on large scales with that
inferred from strong lensing constraints in the cluster core in
order to test the universal NFW profile over three orders of
magnitude in physical scale. The resulting dark matter profile can
also be compared with that determined from the cluster galaxies in
order to determine whether baryons are biased tracers and to
determine the overall mass/light ratio on large scales. 

The cluster Cl0024+1654 ($z$=0.395) was selected as the target for
this purpose. The core had already been imaged with {\it HST} (Smail et
al 1996, 1997a) and the redshift of a multiply-imaged arc
($z$=1.675, Broadhurst et al.\ 2000) tightly constrains the mass
within 200 kpc. Comprehensive spectroscopic samples are available
to large radii from the studies of Czoske et al.\ (2001) and Treu et
al (2003, Paper I). Furthermore, deep multi-band optical and near-infrared
photometry has been made available to aid in tracing the cluster
light. 

A plan of the paper follows. In $\S$2 we review the {\it HST}
observations and the associated ground-based observations pertinent
to this comparative study of the distribution of cluster mass and
light. In $\S$3 we discuss the criteria adopted for selecting the
{\em background galaxies} whose image shapes, corrected for
instrumental distortion, form the basis of the weak lensing
analysis. We test the uncertainties in our measured shear by
comparing shapes derived from our WFPC2 images with a more limited
dataset available from parallel STIS images. We then discuss
various means for locating photometrically the {\em cluster
galaxies} and using their distribution to determine the stellar
mass. The lensing analysis conducted on the data is presented in $\S$4. 
First we examine the 2-D distribution of mass and then discuss 
the subsequent azimuthally-averaged 1-D radial profiles. We
develop a technique for extracting the best fit radial mass
distribution of the dominant cluster component, normalized with
strong lensing constraints in the cluster core. In $\S$5, we compare
the mass and light distribution
over $\simeq$0.1-5 Mpc scales,
and our principal conclusions are summarized in $\S$6.


\section{Observations}

\subsection{Hubble Space Telescope Data}

\begin{figure*}
\centerline{\psfig{file=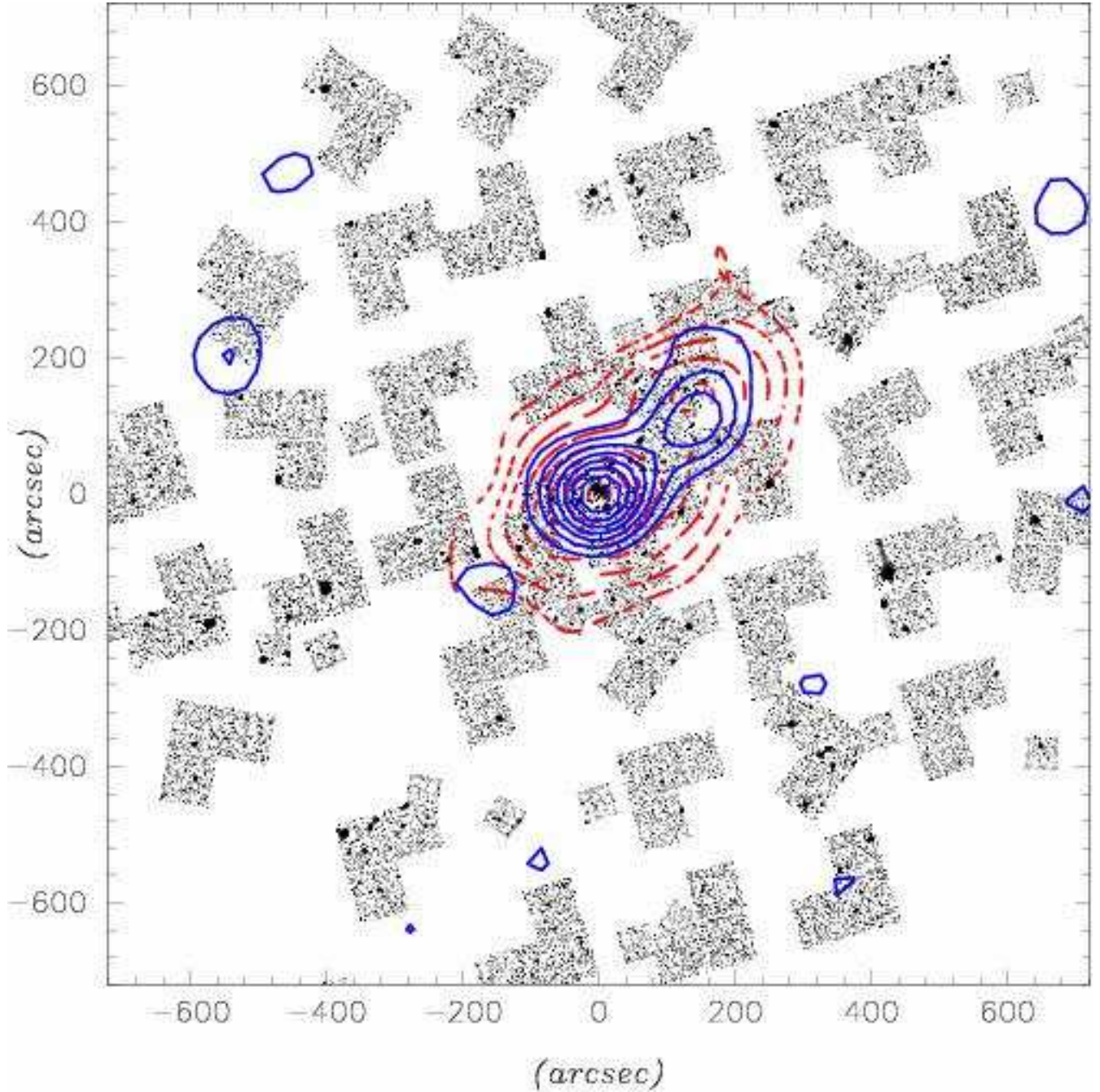}}
\figcaption{The 39 WFPC2/F814W, and the 38 STIS/50CCD pointings
sparsely covering the Cl0024+1654 cluster. The dashed (red) contours
represent the number density of cluster members as derived by
Czoske et al.\ (2002). The solid (blue) contours show the mass map built from
the joint WFPC2/STIS analysis using the {\sc LensEnt} software
(Bridle et al.\ 1998, Marshall et al.\ 2002). North is at the top,
East to the left. The center of this map corresponds to the center of the mass
of the main clump at $\alpha_{J2000} =$ 00h26m35.53s, 
$\delta_{J2000}=$ 17d09m38.0s.}
\end{figure*}

In order to measure reliably the weak lensing signal over the required
range in cluster radius, a large mosaic of 38 independent WFPC2/F814W
and associated STIS/CCD50CLR images was obtained through a substantial
{\it HST} GO campaign (PI: R.  Ellis, ID \# 8559). A fully-sampled
mosaic of WFPC2 images across a 10 $\times$ 10 Mpc field would have
been prohibitively expensive in {\it HST} time. Accordingly, as the
principal goal is to measure the {\em radial} shear profile to
$\simeq$ 5 Mpc, a dilute pattern of pointings was considered
adequate. As discussed in Treu et al.  (2003, Paper I), the original
WFPC2 field centers and orientations were carefully selected, both to
avoid bright stars and to maximize the utility of the independent
parallel STIS images whose overlap with WFPC2 would assist with
calibration of the shear signal. Unfortunately, due to scheduling
constraints the field orientations were later left unconstrained in
order to ensure rapid execution of the program.  The final mosaic is
depicted in Figure~1.  The reduction of the WFPC2 imaging data and
generation of an astrometric solution and photometric catalog are
discussed in considerable detail by Treu et al. (2003, Paper I) to
which the interested reader is referred.  The 38 parallel STIS images,
each covering 51$\times$51 arcsec, were reduced within \textsc{iraf}
using techniques similar to those employed for WFPC2, with the
exception that sub-pixel sampling was not implemented. A photometric
catalog was generated using the SExtractor package (Bertin \& Arnouts
1996) and, to match the STIS characteristics, a detection threshold of
10 contiguous pixels above the 1.5--$\sigma$/pixel isophote was
used. The principle utility of the STIS data lies in verifying, with
somewhat higher fidelity, the weak shear at various points across the
cluster.  Although the total area sampled by STIS is only $\sim 25$
arcmin$^2$ (respectively 171 arcmin$^2$ for WFPC2), the surface
density of background sources is somewhat higher 94 arcmin$^{-2}$
(respectively 83 for WFPC2) and the improved image sampling leads to
more reliable ellipticities within our adopted magnitude limit (see
\S3.1).  The filling factor resulting from our sparse sampled strategy
is 100\% in the inner 50 arcsec, falling to an average of 45\% in the
100-400 arcsec annulus and to 35\% in the 400-700 arcsec annulus.

\subsection{Ground-Based Data}

Ground-based data is used in this study to both locate cluster
members via their broad-band colors and to determine the
associated stellar mass using near-infrared luminosities. Optical
data of Cl0024$+$1654 was taken with the 3.6m Canada 
France Hawaii telescope (CFHT,
see Paper I) using CFHT~12K camera (Cuillandre et al.\ 2000) 
in $B,V,R$ and $I$ (Czoske et al.\ 2001, 2002,
2003). The data reduction is described in Czoske et al.\ (2003).

A series of wide-field near-infrared $K_s$ images of
Cl\,0024$+$1654 were taken on October 29--30 2002 using the
newly-commissioned WIRC-2K 2048$\times$2048 HgCdTe infrared camera
(Eikenberry et al.\ 2002) at the prime focus of the Hale 5.1m. These
data map a $26'\times26'$ area around the cluster center. The
mosaic comprises nine pointings arranged in a $3\times3$ grid,
each encompassing a $8.7'\times8.7'$ field. The total integration
time per grid position was 1.1--ksec. These data were
reduced using standard {\sc iraf} tasks to dark subtract,
linearize, flat field using a local sky median, integer pixel
align and co-add the individual frames to remove defects and
cosmic ray events. The median seeing on the final reduced frames
is FWHM$=(0.93\pm0.10)''$ where the uncertainty represents
the rms variation between fields.

Science observations were interspersed with observations of
standards from the UKIRT list (Hawarden et al.\ 2001). Zero-points
derived from these data do not vary with airmass and time of
observation by more than $\sim0.05$\,mags. We also exploited the
small overlaps between adjacent fields in our $3\times3$ mosaic to
compare independent photometry of objects in the overlap regions.
This exercise confirms that the relative calibration of the 9 WIRC
fields is accurate to $<0.1$mags.

We analyzed our near-infrared mosaic with the SExtractor package
(Bertin \& Arnouts 1996). All objects with isophotal areas in
excess of 5 pixels (0.3 arcsec$^2$) at the
$\mu_{K_S}=21.8 {\rm mag.arcsec}^{-2}$ isophote ($1.5\sigma$
pixel$^{-1}$) were selected. Sources lying close to the
diffraction spikes of bright stars and within $10''$ of the edge
of the field of view were removed. Monte Carlo simulations
determined the completeness limits of these catalogs. Scaled
artificial point source that match the seeing were inserted at
random positions in the $K_S$--band mosaic and examined using the
same SExtractor configuration. The 80\% point source completeness
limit (roughly equivalent to a 5--$\sigma$ detection limit) was
determined to be $K_S(80\%)=19.5$. 

All images ({\it HST} and ground based) were registered to the
astrometry of Czoske et al.\ (2001) and Treu et al.\ (2003)
which is based on the USNO-A2.0 catalog. In this way the
entire photometric (and spectroscopic) dataset is on a
unique and well-defined world coordinate system.


\section{Selection of Galaxy Catalogs}

We now use the data discussed in $\S$2 to construct a robust
sample of {\em background} galaxies whose shapes will permit the
weak lensing analysis, and a list of {\em cluster members} defined
from the ground-based photometry, from which the stellar mass and
cluster light distribution can be determined.

\subsection{Background Sample and PSF Correction}

A master catalog was compiled from the photometric
catalog presented by Treu et al.\ (2003, Paper I) limited at $I$=26,
and the STIS catalogue described in $\S$2.1. To facilitate
comparisons between the STIS and WFPC2 data, we arbitrarily
aligned the photometric zero points of the STIS data (taken with
no filter) so that objects both seen in WFPC2 and STIS have, on
average, similar observed magnitudes.

To construct the background sample catalogue, we first considered
all galaxies with $23<I<26$. The $I$=23 limit corresponds to
$M^{\ast}$+3.7$^m$, more than a magnitude fainter than the point
at which cluster
contamination becomes significant. We then removed: {\it (i)} all
objects closer than 10 pixels from the WFPC2/STIS detector
borders, {\it (ii)} all objects where the adjacent sky background
is 0.5 magnitude brighter than the mode on that WFPC2/STIS chip,
and {\it (iii)} spurious detection near saturated stars. This
removed most spurious detections as well as those contaminated by
a bright object or border effects. Within the adopted magnitude
range $23<I<26$ the total surface densities for the WFPC2 and STIS
photometric catalogs were 48.5 arcmin$^{-2}$ and 58.2
arcmin$^{-2}$ respectively.

\begin{figure*}
\centerline{\psfig{file=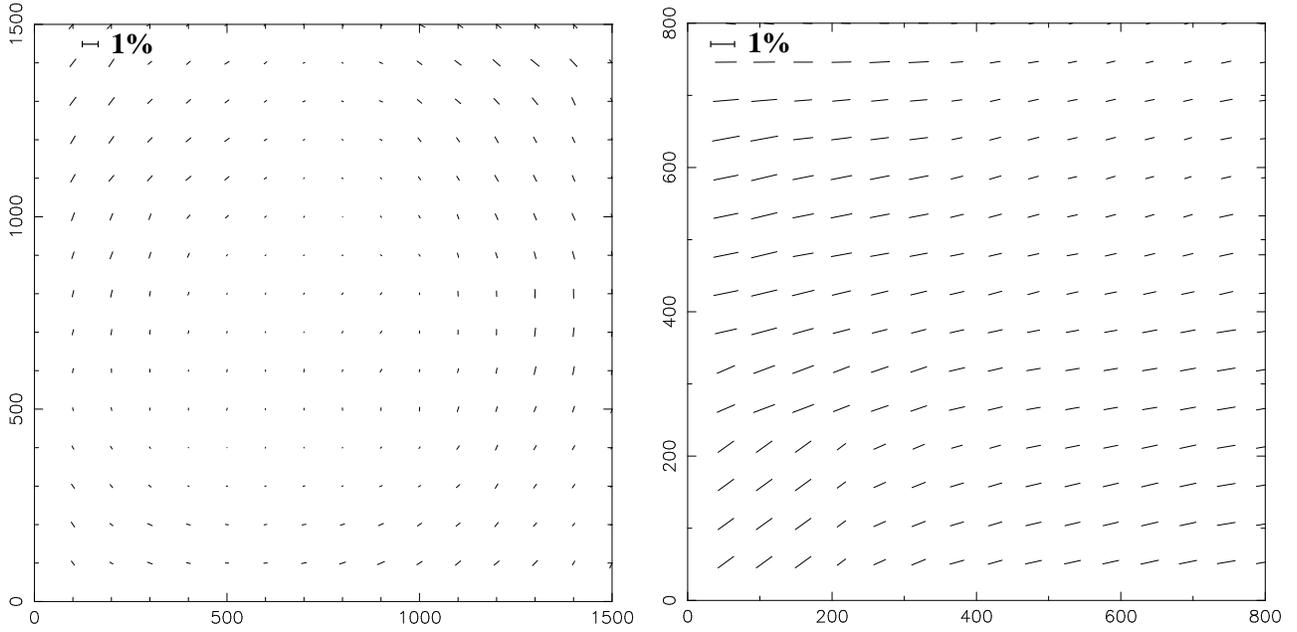}}
\figcaption{Anisotropy of the point spread function as measured on
selected unsaturated stars stacked according to the x-y coordinate
on the relevant detector chip. {\bf Left:} WFPC2 camera distortion
averaged over the three chips. {\bf Right:} STIS camera.}
\end{figure*}

To determine the instrumental PSF distortion and circularization
for the WFPC2 and STIS images, we located and fitted all 
(non saturated) stellar images using a 2-D Gaussian with 
the {\sc im2shape} software developed by Bridle et al.\ (2002, 2003). 
Unlike other direct methods such as that implemented
in {\sc imcat} (Kaiser 2000), {\sc im2shape} is a Bayesian method
in which the galaxy and PSF stars are modeled as a sum of 2
elliptical Gaussians, and the predicted image pixel intensities
are directly compared to those observed as proposed
by Kuijken (1998). {\sc im2shape} 
pays particular attention to the computation of uncertainties on the
measured ellipticities, which can then be taken
into account while investigating the mass distribution. 
In order to evaluate the 
model parameter uncertainties {\sc im2shape} uses the Markov-Chain 
Monte Carlo (MCMC) sampling method ({\it e.g.} MacKay 2001).
{\sc im2shape} has been extensively checked with simulation and
it appears to be very robust (Bridle et al.\ 2003).
In practice, the first step is to determine the shape of the
PSF stars, which then will be used to determine the shape of the
faint galaxies to be used to measure the shear distortion of the cluster.
As there are on average
$\simeq$3.3 stars per WFPC2 chip and $\simeq$1.2 per STIS chip,
stellar images were stacked according to their relative position
on each chip in order to derive the mean distortion map across the
detectors shown in Figure~2. This WFPC2 distortion map agrees well
with that produced independently by 
Rhodes et al.\ (2000), see also Hoekstra et al.\ (1998).

\begin{figure*}
\centerline{\psfig{file=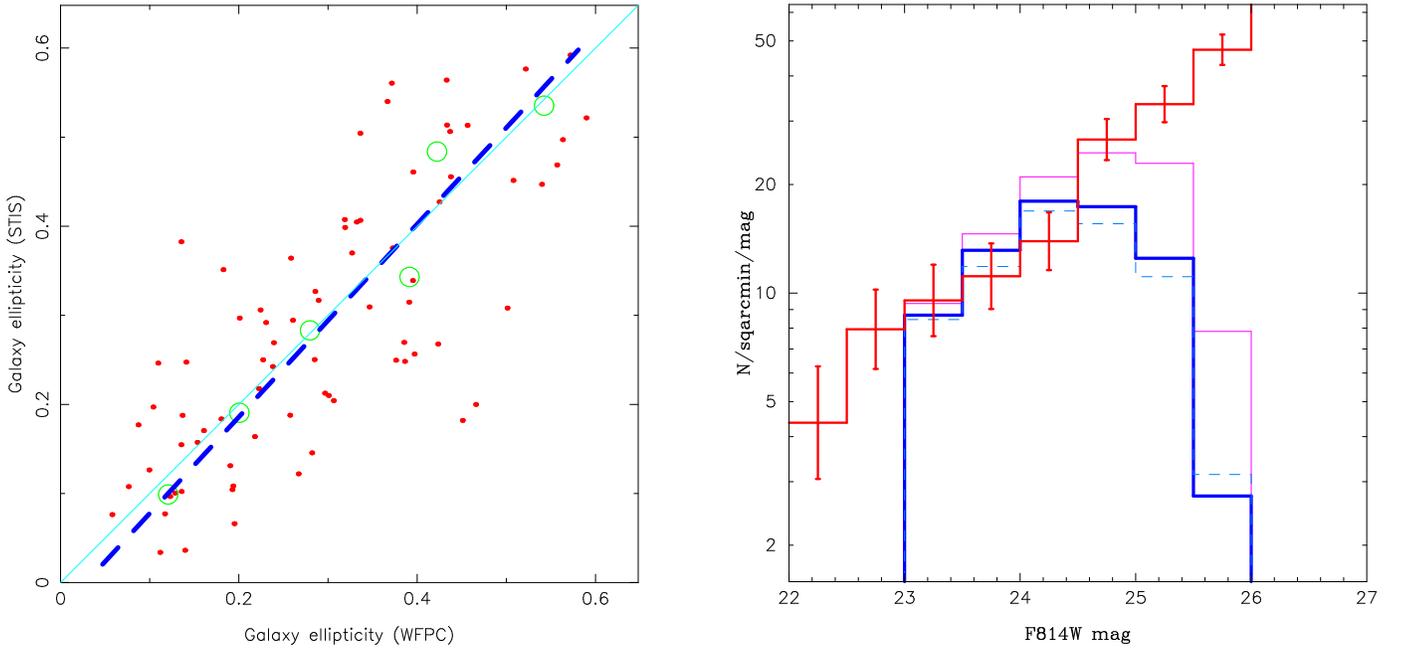}}
\figcaption{
(Left:) Comparison of corrected ellipticity $\epsilon$ measures
for faint background galaxies sampled by both STIS and WFPC2
(restricted to those for which the error $\delta\epsilon$ is
better than 20\%). Circles represent means in bins orthogonal to a
perfect one-to-one relation (thin line). (Right): The extent to
which the background sample can be considered magnitude limited in
the {\it HST} F814W filter. Counts in the HDF South (line
with 1$\sigma$ uncertainties shown) are compared with those in the
Cl0024+1654 photometric catalog (thin solid line) and those selected
for the weak shear analysis ($23<I<$26, $\delta\epsilon<$0.2;
thick=WFPC2+STIS; dashed=WFPC2 only)}
\end{figure*}

The PSF anisotropies derived from stars for both WFPC2 and STIS
data are then used in Bridle et al's {\sc im2shape} package to
correct background galaxies for distortion and circularization.
This correction is applied by fitting each image with a simulated
elliptical galaxy model, of shape $\epsilon =(a-b)/(a+b)$, convolved
with the observed PSF at that point. The ellipticity distribution,
$N(\epsilon)$, required to fit the observed shapes in this manner
has a dispersion $\sigma_{\epsilon} \simeq$0.25. An important
verification of the analysis is a satisfactory comparison of the
corrected ellipticities for those faint 23$<I<$26 galaxies sampled
by both STIS and WFPC2 (Figure~3) and a similar intrinsic distribution for
the background galaxies regardless of the imaging camera used.

An advantageous feature of the {\sc im2shape} software is that it
gives a direct error estimate of each measured ellipticity
$\epsilon$. This is useful in determining which galaxies are used
in the shear measurements. The ellipticity error correlates more
strongly with size and ellipticity than magnitude. The error
distribution is slightly better, as expected, for the STIS data.
Applying a limiting error criterion of $\delta\epsilon<$0.2, we
recover 33.2 galaxies arcmin$^{-2}$ on the WFPC2 data and 45.5
arcmin$^{-2}$ on the STIS data.

By selecting galaxies according to the precision of their measured
shape, the background galaxy catalog departs from a strictly
magnitude limited form and this departure is important in
determining the likely redshift distribution $N(z)$. The extent to
which our selection criteria are equivalent to a pure magnitude limited
can be gauged by comparing the counts in the final catalog with
those derived on blank fields. Noting earlier discussions of a
similar nature in Paper I, for this purpose we chose counts in the
HDF South (Casertano et al.\ 2000, Figure~3). Our shape-restricted
23$<I<$26 sample is broadly equivalent, in effective surface
density, to a $I<$25.2 sample.
In practice, of course, there will be departures between $N(z)$
for our sample and that for a strict $I<$25.2 sample. 
The extent of the departure from a magnitude limited sample
was discussed in the context of photometric redshifts by Hoekstra
et al.\ (2000) who deduced only a 6\% lowering in the derived mean
$\beta=D_{LS}/D_S$ {\it e.g.} that appropriate for a strict magnitude
limited sample, a factor determined by weighting the individual
photometric redshifts according to the associated shape error.
Given there will be uncertainties in the photometric redshifts
which cannot be reliably estimated, we deduce that, at our
effective magnitude limit of $I\simeq$25.2, the mean redshift is
$\overline{z}\simeq 1.15 \pm 0.3$ (based on photometric redshift determination
of the HDFs) where this uncertainty includes
the effects discussed by Hoekstra et al.\ (2000).

\subsection{Cluster Light}

The principal utility of the optical and near-infrared photometry
discussed in $\S$2.2 is to identify galaxies whose
optical-near infrared colors indicate they are cluster
members. Once identified, the
infrared luminosities of these galaxies can be
used to trace the established stellar mass (Brinchmann \& Ellis 2000).

Using the astrometric optical and near infrared catalog,
$(B-K_S)$, $(V-K_S)$, $(R-K_S)$ and $(I-K_S)$ colors were
determined for all sources using a $2''$--diameter aperture on
seeing matched frames. A color-magnitude diagram for various
optical/near-infrared color combinations was inspected and the red
cluster sequence was recognizable in each case. At $z=0.395$, the
redshift of Cl\,0024$+$1654, the 4000--\AA\ break of the evolved
cluster galaxy population lies just red-ward of the observed
$V$--band. We thus expect the $(V-K_S)$ color to be more sensitive
than other optical/near-infrared colors available to us when
selecting likely members (Figure~4).

\begin{figure*}
\centerline{\psfig{file=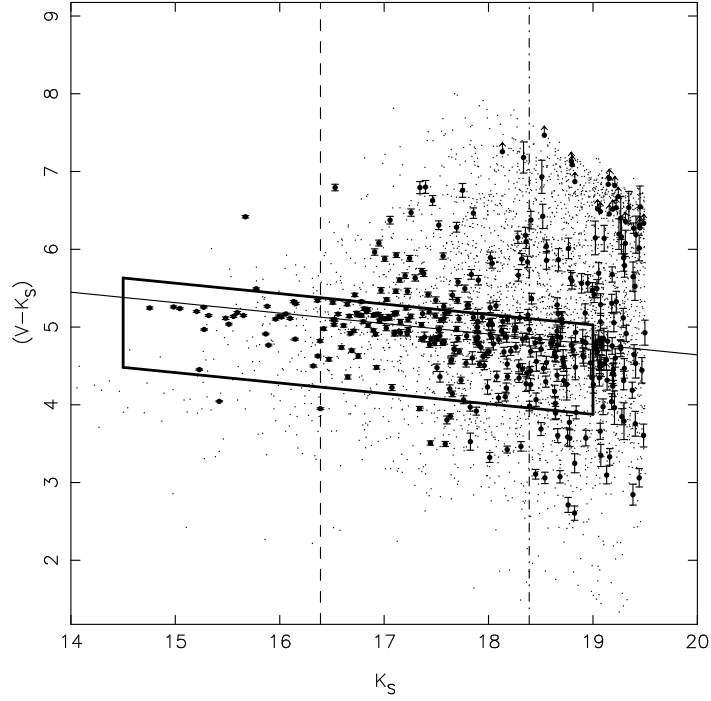,width=0.5\textwidth}}
\figcaption{Color-magnitude diagram ($V-K_s$ vs.  $K_s$) for all
sources in the WIRC-2K mosaic with $K_S<$19.5. Solid symbols
represent confirmed spectroscopic members. The box delineates the
red cluster sequence adopted for determining the distribution of
red cluster light (see text for discussion of its optimal
location). The dashed line indicates the location of L$^*$ galaxy,
and the dot-dashed line L$^*+2^m$. The observed red sequence scatter is 
mainly due to field-field calibration errors, as well as contamination
by dusty edge-on disk galaxies in the cluster core.}
\end{figure*}

To select likely members, we first performed a bi-weight fit to
the $(V-K_S)$ versus $K_S$ color magnitude relation for all
galaxies within a radius of $3'$ of the cluster center down to
$K_S\,<\,K_S^*+2^m$, a region where field contamination is small.
We then used the resulting fit to adopt the selection criterion:
$-0.9<(V-K_S)_{\rm obs}-(V-K_S)_{\rm RCS}<0.25$, where
$(V-K_S)_{\rm obs}$ is the observed color of a galaxy and
$(V-K_S)_{\rm RCS}$ is the color of the red cluster sequence at
the $K_S$--band magnitude of each galaxy obtained from the fit
above. We also applied a cut in $K_S$--band magnitude of $14.5\le
K_S\le19$. Galaxies redder than this selection are likely at
higher redshifts and bluer objects will be foreground galaxies or
star-forming cluster members.

The limits $\Delta(V-K_S)_{\rm red}=0.25$mag and
$\Delta(V-K_S)_{\rm blue}=0.9$mag were chosen after reference to
the extensive spectroscopic catalog (Paper I). In selecting these
limits we considered the ``completeness'', which we defined as the
fraction of spectroscopically-confirmed cluster members that are
photometrically selected by a given set of color-selection
criteria, and the ``purity'', defined as the fraction of
photometrically-selected galaxies that have been spectroscopically
identified and lie within $|\Delta z|<0.05$ of the nominal cluster
redshift. These quantities are plotted as a function of the
location of the red and blue cuts in Figure~5 where the greater
sensitivity of the blue cut is evident. Our final color cuts are
consistent with an $80\%$ completeness limit and a purity level of
approximately $70\%$. We also tested our adoption of $(V-K_S)$ as
the optimum color for this purpose by repeating these experiments
using the $(B-K_S)$, $(R-K_S)$ and $(I-K_S)$ data. The results are
very similar in all four colors. However at a fixed completeness
of $80\%$, as expected, the $(V-K_S)$--based catalog suffers less
contamination (i.e.\ higher purity) than the other colors by
$\sim10$--15\%.

Field contamination in both the $K$-limited sample, and the {\em red
sequence} sample additionally constrained by $V-K$ color, can be
determined from the source density in independent offset fields
available to us (K.Bundy priv. commun.). However, this leads to an
over-correction for the periphery because, as discussed in Paper I,
Cl0024+1654 is surrounded by a region of density slightly lower than
average. To avoid a negative density, we therefore applied an
{\em in situ} background correction using the source density measured
in the 4$<r<$5 Mpc annulus around Cl0024.  The statistical uncertainty
associated with this correction, and the bias associated with members
lying within this outer annulus, is included in the subsequent
analysis.

\begin{figure*}
\centerline{\psfig{file=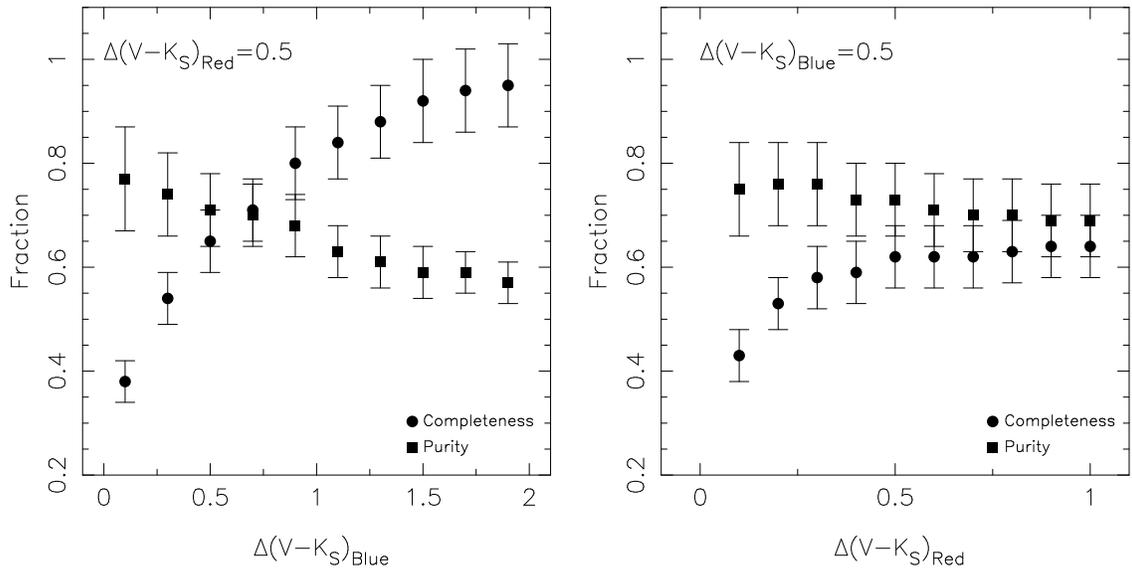}}
\figcaption{ The purity and completeness membership fractions (as
defined in the text) as a function of the location of the blue (left)
and red (right) $V-K_S$ color envelope.}
\end{figure*}

From the $K_s$-limited and red sequence galaxy catalogs, we
computed projected luminosity maps for both the raw $K$-band
light, limited to 2400 galaxies brighter than $K\simeq$19, and
for $\simeq$500 red sequence objects selected by the
procedure discussed above (Figure~6). Both maps are adaptively
smoothed by selecting areas containing 15 suitable galaxies and
dividing their integrated light by the corresponding area.
Although the mean smoothing radius in the peripheral regions is
$\simeq$200 arcsec (1.15 Mpc at $z$=0.4), in the core where the
density is higher, the smoothing scale approaches $\simeq$30
arcsec ($\simeq$170 kpc).

This comparison should give a good indication of the effects of
projection and how to distinguish these from genuine substructure
in the cluster. The raw $K_s$ image should be sensitive to
structures projected over all redshift whereas the latter should
preferentially locate stellar mass broadly within the redshift
range 0.3-0.5.  The most conspicuous feature in both of these
distributions (other than the cluster core) is a secondary
concentration of light $\sim$3 arcmin to the NW of center, coincident
with the secondary concentration of spectroscopically confirmed
cluster members shown in Paper I. The $K_s$ band luminosity ratio
of the two clumps, measured within a 0.5 Mpc diameter circular
aperture is 3.6:1. We will continue the discussion of this
secondary clump in $\S$5, in the context of our mass model. The
periphery of the cluster does reveal minor fluctuations in the
$K_s$ image but as color selection satisfactory removes these, it
would seem the bulk of the stellar mass in Cl0024+1654 is reasonably
smooth and well-behaved.

\begin{figure*}
\centerline{\psfig{file=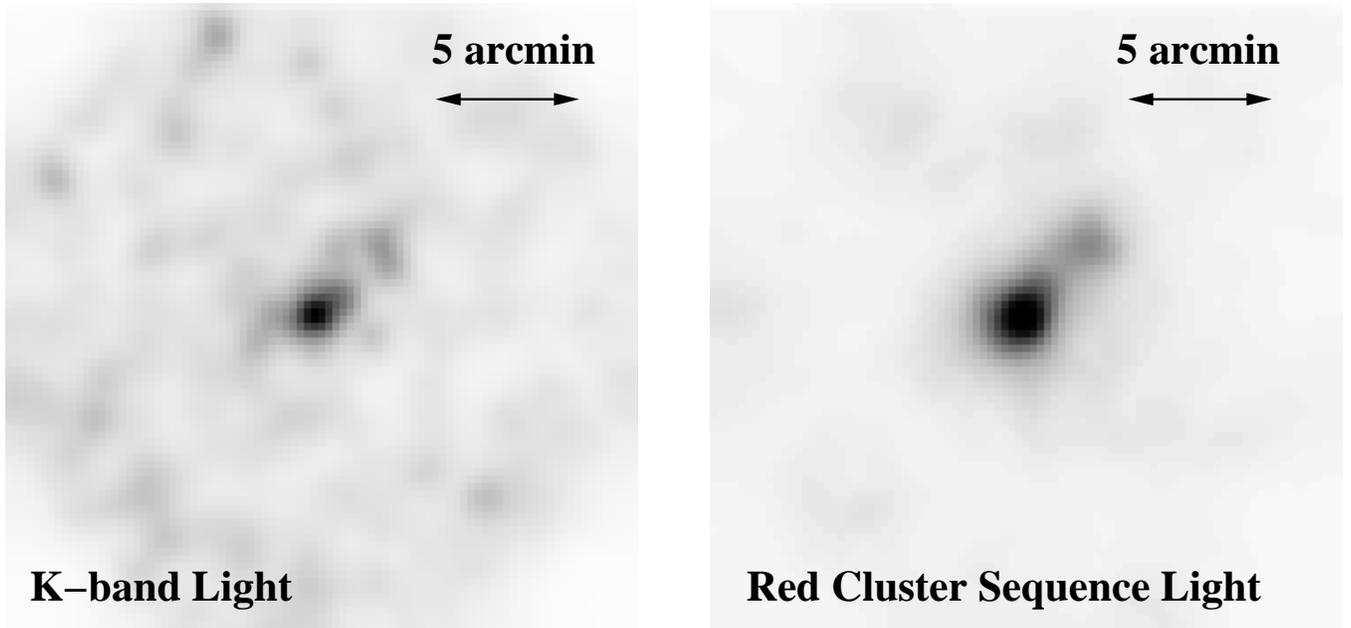}}
\figcaption { The distribution of stellar light as determined from the
field-subtracted $K_S$ photometry (left) and that for red sequence
galaxies selected according to $V-K_S$ color (see text, right). The
secondary substructure originally located by Czoske et al (2002)
is evident in both panels NE of the cluster
core which matches that seen in the weak lensing map (Figure~1). }
\end{figure*}


\section{Lensing Analyses}

\subsection{Two-Dimensional Mass Distribution}

Our goal is to characterize the radial dependence of total mass
for Cl0024+1654 in the context of popular theoretical realizations
and to compare the mass profile with that determined for the
stellar light. Figure~6 suggests that cluster substructure may
complicate our analysis. Accordingly we proceed as
follows. We first apply a non-parametric method to derive a
projected 2-D mass map from our weak shear data. Although our
WFPC2 sampling strategy was designed to yield a 1-D mass profile
in the cluster periphery, to an intermediate radius, the
signal/noise is adequate for a 2D mass reconstruction. 
We then explore a multi-component weak lensing analysis
where we then incorporate strong lensing constraints
derived from multiple images in the cluster core.

Since the gravitational shear is not a local measure of the
projected mass density, deriving a mass distribution requires an
inversion process (Mellier 1999). From the distortion-corrected
samples discussed in $\S3.1$ we computed the associated projected
mass map using the maximum entropy {\sl LensEnt} method discussed in
Bridle et al.\ (1999, see also Marshall et al.\ 2002). In order to
maintain more than 20 galaxies in each (non-empty) cell, we
defined a reconstruction grid of 64$\times$64 cells, each 30
arcsec on a side. With this parameterisation, we infer from the 
data an intrinsic correlation function width,
which reflects the size of the smallest detectable feature on the
mass map, is $\simeq$100 arcsec (570 kpc).

The resulting 2-D mass distribution is overplotted as a contour
map on Figure~1. It bears a striking resemblance to that already
derived for the infrared and red sequence light (Figure~6). In
addition to resolving the main cluster core which is roughly centered on
the peak of X-ray emission determined from the recent {\it Chandra}
image (Tozzi et al., in preparation), a weaker second mass peak is
detected with 10$\sigma$ significance $\simeq$1 Mpc NW of the
cluster center (this substructure has already been identified in the work of
Czoske et al.\ 2002). The location of this peak closely matches that
seen in Figure~6. Moreover its relative prominence to the main
clump is similar to that deduced from the $K$-band light; indeed within a
0.5 Mpc diameter, it corresponds to $\simeq$25-30\% of the mass of the main
clump in the same aperture. Thus the two mass clumps have similar M/L ratio.
Other features in Figure~1 have a significance of less than 3$\sigma$.

\subsection{Weak Shear Profile}

The 2-D mass distribution becomes very noisy beyond a radius of
$\simeq$400 arcsec (2.3 Mpc) and to make further progress it is necessary to
azimuthally average the shear and consider only the 1-D profile.
In order to derive a radial profile, it is first necessary to
pinpoint precisely the cluster center. From the 2D weak lensing
mass map and the strong lensing model of the multiple image system
in this cluster ($\S$4.4) we determine the center of mass to be
at: $\alpha_{J2000} =$ 00h26m35.53s, $\delta_{J2000} =$
17d09m38.0s. The brightest cluster galaxy lies only $\simeq$5
arcsec N of this location. Our adopted center is also within 5
arcsec SW of the peak of {\it Chandra} emission (superimposed on a luminous
cluster galaxy).

Galaxy shapes were azimuthally-averaged within circular bins
within fixed radial intervals. Figure~7 shows the resulting radial
shear profile as measured from the STIS+WFPC2 mean tangential
ellipticity $<\epsilon_{tan}>$. A smooth trend is seen in both the
overall trend and the dispersion, suggesting that the secondary
substructure identified in Figures~1 and~6 has only a minor effect
on the overall profile.

Cluster shear has not hitherto been detected to radii where the
shear is $\lessapprox$1\%, so it is important to demonstrate the
reliability of our detections. Figure~7 compares the outermost
signals in the two final radial bins from 400$<r<$800 arcsec as
measured independently using WFPC2 and STIS data alone which are seen to
agree. In combination the shear within these two bins is
significant at the 4$\sigma$ level.
Figure~7 also includes the absolute value of the radial shear 
(equivalent to the standard 45 degree test).
In the case of ideal data with one single mass component, 
the reduced radial shear profile should average to zero, with statistical 
errors inversely proportional to the square root of the number galaxies 
used to compute the shear.  
The absolute value of the reduced radial shear thus provides an 
upper limit to the systematic error on the reduced tangential shear and the
contribution from additional mass clumps. Figure~7 summarizes the results:\\
1) except for the last bin of the WFPC2 data,
          the tangential shear is well above the radial shear,
	demonstrating minimal interference from systematic errors.\\
2) the STIS data have a lower systematic error compared to the WFPC2
          data, as expected.\\

\begin{figure*}
\centerline{\psfig{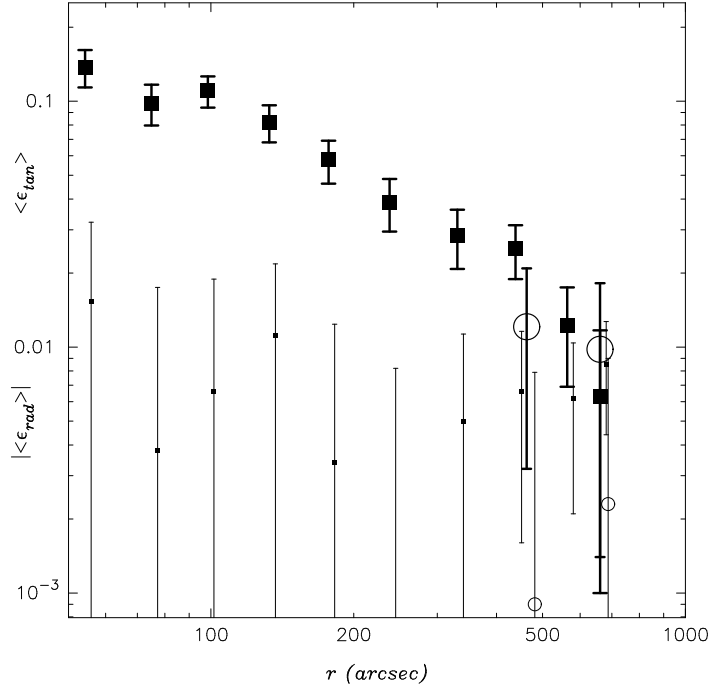}}
\figcaption{
The thick symbols show the reduced tangential shear, azimuthally-averaged, for
the WFPC2 and STIS sample with 1-$\sigma$ uncertainties derived
by comparing different sectors.  The significance of the shear
detected in the outermost radial bins can be evaluated by examining
the STIS measurement alone (open circles with larger error bars).
The thin symbols show the absolute reduced radial shear 
(equivalent to plotting the tangential shear of the galaxies
for which the orientation would have been changed by 45 degrees). In all
cases except the last WFPC2 point, the absolute reduced radial shear
is well below the reduced tangential shear indicating that systematic
error are not important. Note that for STIS data, the radial component
is always well below the tangential component.
}
\end{figure*}

These measurements can be compared to previous weak lensing analysis of
this cluster.  In a truly pioneering paper, 
Bonnet et al.\ (1994) reported on the
detection of shear to a radius $\lesssim$2.3 Mpc from the center of
Cl0024+1654 using panoramic imaging taken with the Canada France
Hawaii Telescope.  The shear observed by {\it HST} at $\simeq$2
Mpc (350 arcsec) is $\simeq$2\% which is half 
the value quoted in Bonnet et al.\ (1994).
The discrepancy most likely lies in over-correction of the ground-based
PSF circularization applied by Bonnet et al.\ (1994)
which was done at a time where PSF corrections
were just being experimented.

\subsection{Deriving Mass Profiles: Weak Shear}

We face two complications in deriving a self-consistent radial
mass profile. First, within a certain cluster radius,
contamination of the background sample selected with 23$<I<$26 by
cluster galaxies will increasingly dilute the shear signal. This
can be seen via the flattened slope of the shear profile at radii
$r<$100 arcsec (575 kpc) in Figure~7. Secondly, cluster substructures will
deflect the radial profile from that appropriate to fitting a
single component. Fortunately, both effects are straightforward to
account for.

In the case of shear dilution from cluster members, the radial
surface density of background galaxies gives a good measure of the
likely contamination as a true background population should show
no strong radial variation. Assuming that the contaminating
population has random orientation and an ellipticity distribution
similar to that of the genuine background population, we can
easily correct the measured shear via:

\begin{equation}
<\epsilon_{tan}>_{corr}(r)= \frac{n}{n(r)} <\epsilon_{tan}>(r)
\end{equation}

where $n$ is the mean number density of galaxies in our background
galaxy catalogue and $n(r)$ represents its variation as a function
of cluster radius.

Concerning the effect of a major substructure, such as the
secondary mass concentration $\simeq$3 arcmin NW of the center
detected in Figures~1 and~6, we choose to permit new components in
addition to the main cluster mass distribution in our 2-D model
fitting, assuming (for convenience) all share similar mass
profiles whose universal form we consider to be the wanted
unknown. We then compute analytically the
reduced shear field~$g=\gamma/(1-\kappa)$ arising from a linear combination 
(by mass) of such clumps.

Assuming a Gaussian intrinsic ellipticity distribution and
Gaussian image shape measurement errors the log-likelihood becomes:

\begin{equation}
\log{\pr\,(\data|\pars,N)} =  \log{\frac{1}{(2 \pi \sigma_{\epsilon}^2)^{N}}}
         - \sum \frac{|\epsilon - g|^2}{2 \sigma_{\epsilon}^2}
\end{equation}

\noindent where $N$ is the number of clump fitted, and
 $\sigma_{\epsilon}^2 = \sigma_{shape}^2 +
\sigma_{intrinsic}^2$ and $\pars$ are the sets of parameters for
the profile of each clump.

We now turn to determine the most likely mass profile based on the
weak shear data. Our analysis is based on two physically-motivated
profiles which we fit to the Cl0024+1654 data. First, we
consider the universal profile predicted by hierarchical
clustering models (NFW). The dark matter density
$\rho$ is described by:

\begin{equation}
\frac{\rho(r)}{\rho_{\rm crit}}=\frac{\delta_c}{(r/r_s)(1+r/r_s)^2},
\label{eq:NFW}
\end{equation}
\noindent
where $r_s$ is a scale radius, $\delta_c$ is a characteristic 
(dimensionless) density, and $\rho_{crit} = 3H_0^2/8\pi G$ is 
the critical density for closure. Those parameters can be expressed
in terms of M$_{200}$ and $c$ (NFW). The former 
is defined as the mass enclosed within $r_{200}$ 
(the radius of the sphere enclosing an average density $200\rho_{\rm
crit}$) and gives
a characteristic cluster scale mass which is popular with numerical 
simulations.
The latter is the concentration parameter defined as $c=r_{200}/r_s$.

Second, we consider a Singular Isothermal Sphere (SIS):
\begin{equation}
\rho(r)=\frac{\sigma_{SIS}}{2\pi G r^2}.
\label{eq:SIS}
\end{equation}

The SIS is not only a good approximation of the total mass density
profile of spiral (van Albada \& Sancisi 1986) and early-type
galaxies (Treu \& Koopmans 2002) but has also been justified as
physically meaningful for the outcome of the gravitational
collapse of a system under certain conditions (Fillmore \&
Goldreich 1984).

In addition to these 2 physically motivated models, we 
also consider the more general question of the form of the
mass distribution of large scales by considering a `cored power
law' (CPL) fit with projected mass distribution:
\begin{equation}
\rho(r) = \rho_0 \frac{ 1+ \frac{2}{3}\alpha s^2}
		{(1+s^2)^{5/2-\alpha}}
\end{equation}
where $3-2\alpha$ is the slope of the density mass profile at large radius,
and $s=r/r_c$ with $r_c$ the core radius.
The mass profile of a cored isothermal sphere (CIS) 
corresponds to $\alpha=1/2$.

Our rationale is to determine, for an adopted mass profile,
the likelihood of substructure as revealed by the weak lensing
data. To explore the effect of mass concentrations, other than the
main cluster, we permitted the centroids to vary with uniform
prior probability across the full angular range of our data. A
Markov-Chain Monte-Carlo algorithm ({\it e.g.}
Gilks, Richardson \& Spiegelhalter 1996) was
used to draw samples from the resultant posterior density
$Pr(\data|\pars,N)$. This technique allows efficient characterization 
of this distribution, lending itself to straightforward marginalization 
permitting realistic and accurate estimation of parameter uncertainties.

Figure~8 shows the result of this multi-component shear fitting
technique in the context of an assumed NFW profile. As expected
from Figure~1, only two significant components are necessary to
match the data and the second clump contains about 30\% of the
total mass, as measured via $M_{200}$ (Figure~8). The Bayesian
evidence $\pr\,(\data|N)$ applied with a uniform prior on $N$ in
the range 0:2 gives the ratio

\begin{equation}
  \frac{\pr\,(N=2|\data,\textrm{NFW})}{\pr\,(N=1|\data,\textrm{NFW})} =  25.
\end{equation}

\noindent which acts as a measure of the significance of substructure (i.e.
a two-clump model is a 25 times more likely representation of the
data than a single component). The evidence for a three clump
model is no greater within the numerical precision.

\begin{figure*}
\centerline{\psfig{file=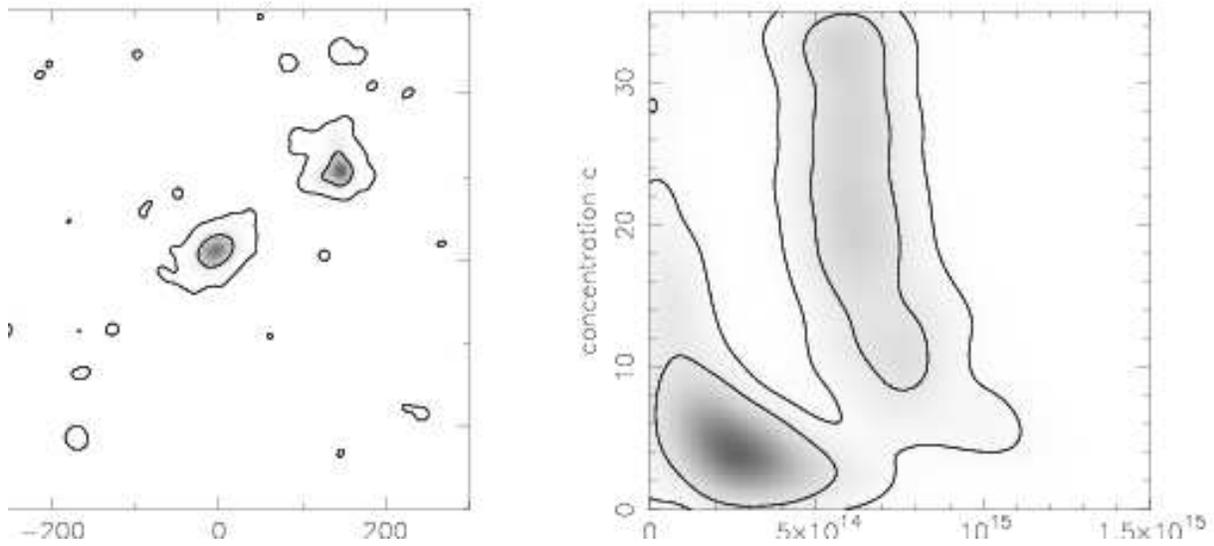,width=0.9\textwidth}}
\figcaption{ Weak lensing constraints on the model fitting
parameters. (left) Marginalized posterior probability distribution
$\textrm{Pr}(x,y|\textrm{\,Data, NFW\,})$
for the multiple component lens model assuming a NFW
profile which accurately locates the positions of no more than two
clumps.
(right) Marginalized posterior probability distribution
$\textrm{Pr}(M_{200},c|\textrm{\,Data, NFW\,})$
for a two component lens model showing
the NFW profile parameters for both components.
 }
\end{figure*}

The above analysis therefore permits us to accommodate, in a
fairly rigorous way, substructure in Cl0024+1654 within the context
of a universal density profile. We now turn our attention to the
key question of determining the most accurate form for this
universal profile, fixing the number of components in the lens
model to $N=2$. 
The probabilities for the different models (CPL, SIS, CIS)
relative to the NFW model are summarized in Table~2.
Both the SIS and CIS model were found to be less likely given our
weak lensing data than the CPL and NFW models.

To underline the importance of the radial extent of the data, we
repeated the above analysis using only galaxies within 420 arcsec from
the cluster centre. We find the ratio $Pr(NFW)/Pr(SIS)$ to be
$\sim$2.5 (as opposed to 15 for the full dataset; Table~2).  The
respective model parameter inferences are unaffected, and their
derived uncertainties only marginally larger. We thus conclude that
the full dynamical range is needed to distinguish between an NFW and
an SIS mass density profile.

Due to sparse sampling,
at radii larger than 200~arcsec it would be easy to miss a small
mass concentration lying in the area not covered by the HST data.
We investigate a three clump model, where the 2 first clump priors are set
to the posteriors of the 2 clump model. The position of the third clump 
was allowed to be anywhere within 600 arcsec from the center. 
We found that to retain
the goodness of fit, the third clump would have to have $M_{200}$ 
smaller than $1.5\,10^{14}$ M$_\odot$ at 90\% confidence (this mass
correspond to half 
the mass of the second clump). Furthermore, adding a third clump with
such a low mass was found not to affect significantly the best fitting
parameters of the 2 main clumps.
Beyond 600~arcsec, the coverage is too sparse to  be able to put 
meaningful constraints on any putative mass clumps-- any such objects 
would indeed remain either undetected or confused with noise.
We conclude that apart the 2 main clumps,
there is no other significant massive clump that can be detected by the data.
Would such clump exists, its low mass would not change our conclusions.

\subsection{Deriving Mass Profiles: Incorporating Strong Lensing}

We now combine constraints on the outer ($>$100 arcsec - 575kpc) mass
profile from weak lensing signals with those available from strong
lensing constraints (Smail et al.\ 1996) based on modeling the location of
the multiply-imaged arc whose spectroscopic redshift is known
($z$=1.675, Broadhurst et al.\ 2000).

The multiply-imaged arcs are on average 27 arcsec (155 kpc)  
from the cluster center and
enables us to place a tight prior on the Einstein radius of the
central clump. A second clump of the same circularly-symmetric NFW
profile is added, and the posterior probability distribution of
the parameters investigated. We constrained the location of the
central clump ($\pm$1 arcsec) and permitted the second clump to
lie anywhere within a radius of 5 arcmin from the cluster center.

The evidence can be used as a powerful tool to probe the
consistency of the results arising from the weak and strong
lensing datasets by comparing that determined here with that
inferred for the same model parameters from the weak lensing data
alone. The tight strong lensing prior reduces the volume of the
parameter space in the region of high likelihood, making the weak
lensing data more probable, indicating very good consistency
between the two datasets.

Figure~9 shows the two-dimensional probability distributions of
the concentration and $M_{200}$ for the NFW model. Note
that including the strong lensing information improves both the mass
and concentration precision for \emph{both} clumps (by comparison of
Figure~8 and~9). 
The NFW concentration parameter
of the main clump is of particular interest. Numerically-simulated
clusters typically have concentrations of around 5, somewhat
smaller than that inferred for the main clump of Cl0024+1654 (Table
1).

The probabilities for the different models (CPL, SIS, CIS)
relative to the NFW model are summarized in Table~2.
Both the SIS and CIS model appear to be less probable given 
the data by a factor of some tens of thousands.
Compared with the NFW profile, the CPL has a slightly greater
likelihood, but this advantage is partially a result of the added
flexibility arising from the extra parameter. Indeed, the evidence values
computed for each model are identical within the uncertainties; we
consider there to be no reason to reject the NFW fit.
However, the CPL provides an estimate of the slope at large radii:
we calculate a 95\% bound of $\alpha < 0.3$ (for the main clump)
which translates to an asymptotic
logarithmic slopes of the 3-D density distribution 
of $n > 2.4$ where $\rho(r) \sim r^{-n}$ as
$r\rightarrow\infty$.

\begin{figure*}
\centerline{\psfig{file=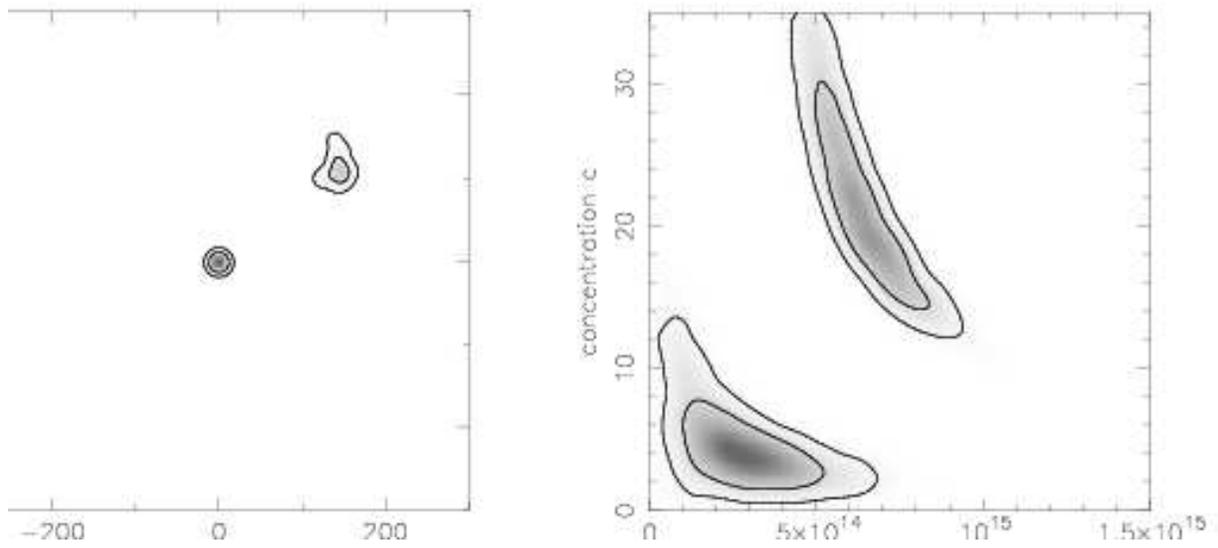,width=0.9\textwidth}}
\figcaption{ Same as Figure~9 when adding in the strong lensing
constraints.}
\end{figure*}

\begin{figure*}
\centerline{\psfig{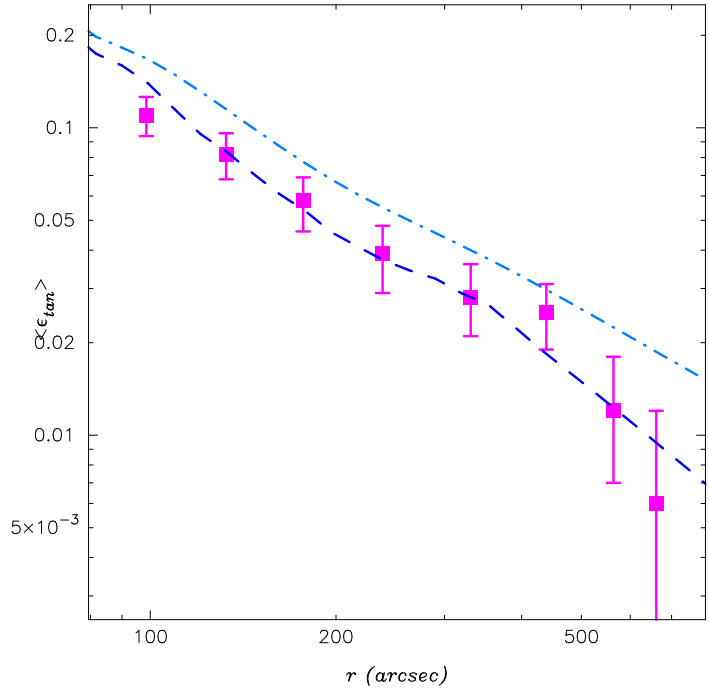}}
\figcaption{ 
Reduced tangential shear profile for the
combined WFPC2 and STIS data (magenta points with error bars).  
The dashed line is the visual representation of
the reduced tangential shear of the 2 clump
NFW model that best fit both the strong and weak lensing constraints.
The dot-dashed line corresponds to
the reduced tangential shear of the 2 clump
SIS model that best  fit the strong lensing constraints but fails
to fit the weak lensing measurements.
}
\end{figure*}

Figure~10 summarizes the above analysis in the context of the
observed reduced tangential shear profile, reproduced using the
combined WFPC2+STIS data from Figure~7. The NFW fit for the two
components in Table 1 accurately represent the data whereas the
isothermal model is clearly discrepant and can not account for
the low shear value at large radius as well as fitting the 
strong lensing constraints.

The only way to fit both the strong lensing and weak shear constraints
with a SIS model would be to significantly modify the redshift
distribution of the background galaxies. Specifically, in order to
satisfy both set of constraints with an SIS, we need to increase the
critical density by 40\%.  This corresponds to shifting the source
plane to a mean redshift of 0.72 inconsistent with our knowledge of
the redshift distribution of faint galaxies.

Finally, following the techniques demonstrated in Kneib et al.\
(1996), we investigated the detailed mass distribution of the inner
cluster including the contribution from galaxy halos (Figure~11).  The
mass distribution arising from galaxy halos were determined using
scaling relations discussed in Natarajan et al.\ (1998) based on the
$V-K_s$-selected sample discussed in $\S$3.2.  Within these
assumptions both the NFW and CPL 2-clump mass models were
investigated. In both cases the exact position and shape of the
5-image multiple can be reproduced, with a slight preference for the
CPL model. The contribution by mass from galaxy halos is similar to
that found in previous cluster analyses (Natarajan et al. 1998,
2002). The results and the variation of this galaxy halo components
with radius will be discussed in a further paper in the series
(Natarajan et al. 2003).

\begin{figure*}
\centerline{\psfig{file=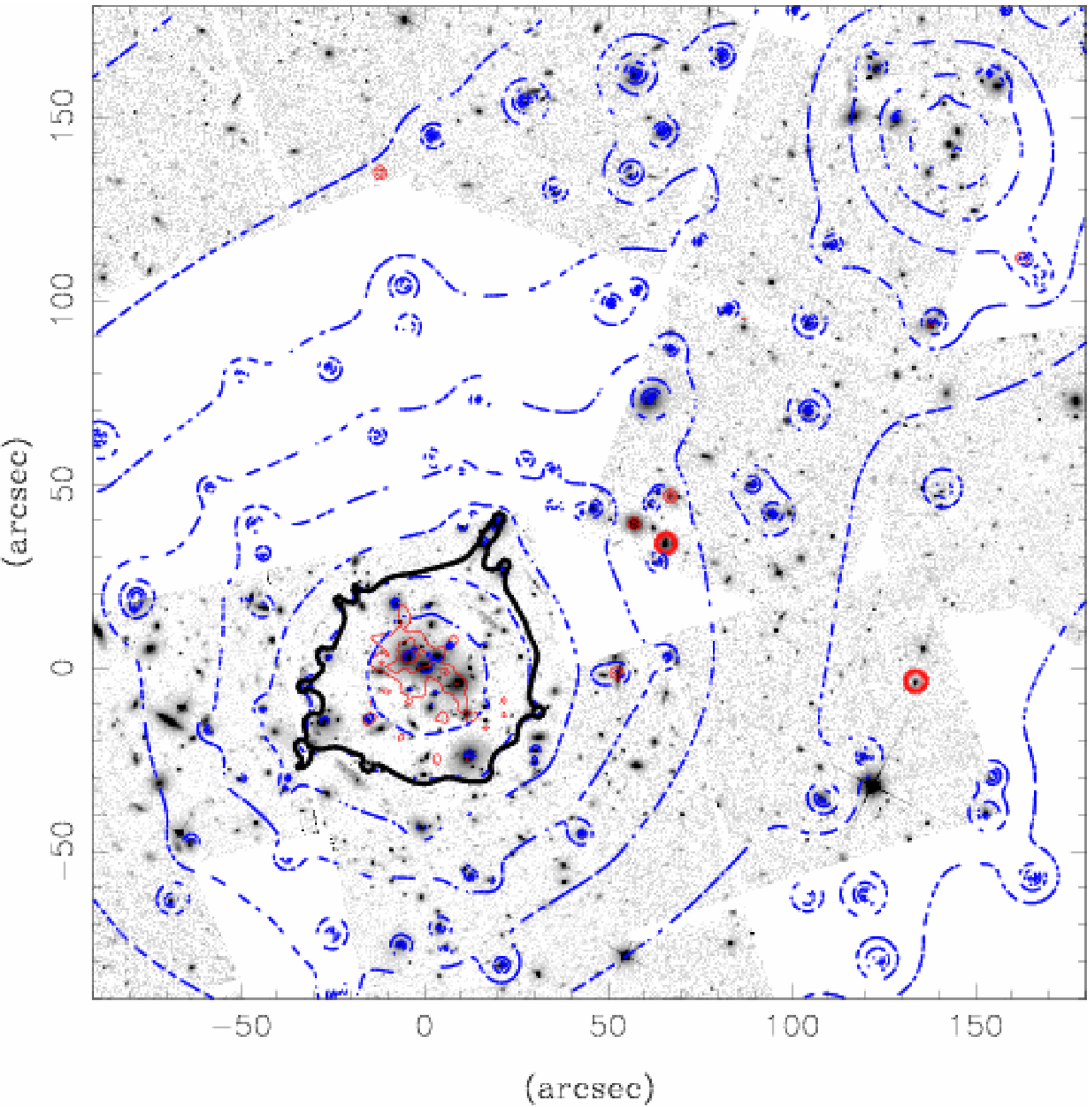}}
\figcaption{Mass distribution in the central part of Cl0024+1654 (dash-dotted
 blue contours). 
The solid (red) contours represent those of {\it Chandra} X-ray emission.
The critical line at the multiple
image redshift (z=1.675) is the thickest (black) line.
The orientation and center are similar to those in Figure~1.  }
\end{figure*}

\section{Comparing Mass and Light}

Given the similarity of the substructure revealed in Cl0024+1654
from the distribution of infrared light (Figure~6) and that
inferred by weak lensing (Figures~1 and~11), we
now compare the overall radial profiles of mass and light. This is
important to understand the degree to which cluster masses
might be under/over-estimated, for example if the DM profile was 
assumed to follow an SIS model.  Our unique data 
gives the first indications, based on lensing, of the extent to
which mass and light trace each other on scales above 1 Mpc.

The integrated mass profile $M(<r)$ derived from the NFW
components listed in Table 1 can be compared with that of stellar
light from the $K_S$-limited and red-sequence maps presented in
Figure~6. The enclosed total mass from the lensing analysis and
the stellar mass determined from the field-corrected $K_S$-limited
catalog ($\S$3.2) track each other very closely (Figure~12).

The extent to which the mass/light ratio of the enclosed
population might vary with radius is examined for both the
$K_S$-limited and red sequence populations in Figure~13. Here, the
uncertainties represent those associated with both background
subtraction and sampling statistics. Given the uncertainties,
there is little convincing evidence for any segregation of mass
and light. 
The mean $M/L_K$ (restframe corrected) for the red sequence galaxies is
$\simeq$40 and that for the overall cluster galaxies $35\pm5$ 
$M_\odot/L_\odot$ (at the $r_{200}$ radius).

Assuming that the total luminosity of the cluster evolves
according to passive evolution of an old stellar
population\footnote{Computed using a 8 and 12 Gyrs single stellar
population synthetic spectrum from Bruzual \& Charlot 1993,
GISSEL96 version, with Salpeter IMF.}, this corresponds to
$\approx42\pm6$ $M_\odot/L_\odot$ ($48\pm6$ for the red galaxies) at $z=$0. 
For comparison, $M/L_K$ at large radii in the Coma cluster is found to
be $49\pm15$ $M_\odot/L_\odot$ from dynamical analysis 
(Geller, Diaferio \& Kurtz 1999; Rines et al.\ 2001). 
Thus passive evolution of an old
stellar populations appears to be consistent with the evolution of
the cluster light as a whole ({\it e.g.} Hoekstra et al.\ 2002). A
similar conclusion follows from the mass to light ratio derived in
the observed $I-$band ($M/L_V$=165$\pm 15$ $M_\odot/L_\odot$) which compares
favorably with previous weak lensing studies of the central
regions (1 Mpc) of clusters at intermediate redshift ({\it e.g.}
Lombardi et al.\ 2000; Hoekstra et al.\ 2002).

A constant mass-to-light ratio out to large radii was determined by
Carlberg, Yee \& Ellingson (1997) using dynamical analysis of an
ensemble cluster from the CNOC dataset (see also van der Marel et
al.\ 2000). Our result extends and reinforces this result given
that our lensing analyses are independent of the dynamical state
of the cluster, do not involve assumptions on the orbital
properties of members nor on the dynamical equilibrium of
infalling galaxies ({\it e.g.} discussion in Biviano \& Girardi 2003).
Furthermore, obtaining this result for a single cluster (as
opposed to an ensemble cluster) ensures that the constancy of
$M/L$ is not an artifact of the scaling laws adopted in
constructing the ensemble.

From ground based data using the WFI 2.2m telescope, Clowe \&
Schneider (2001, 2002) studied the mass profile of three massive
clusters out to $\lesssim 3 h_{65}^{-1}$. Based on a weak lensing
analysis only, they found that they cannot clearly distinguish between
NFW and isothermal; however, in the case of Abell 1689, they did not
reach any good agreement between the strong and weak lensing
constraints.  
In a recent paper, Gavazzi et al.\ (2003)
investigate the mass distribution around the cluster MS2137-03 (see
also Sand et al.\ 2002). They combine strong and weak lensing
measurements up to 900$h_{65}^{-1}$ kpc.  
Although their methodology is similar to ours, their data do not
discriminate between the isothermal and NFW slopes at large radii.

An extension of such weak lensing analysis to the field was conducted
using ground-based data by Wilson et al.\ (2001). They correlated the
2-D distribution of $V-I$ selected red galaxies with shear-based mass
reconstructions for 6 blank fields. Their azimuthally-averaged
luminosity auto-correlation and mass/luminosity cross-correlations
lead them to infer similar profiles for mass and light on scales
$\simeq$0.3-1 Mpc.

\begin{figure*}
\centerline{\psfig{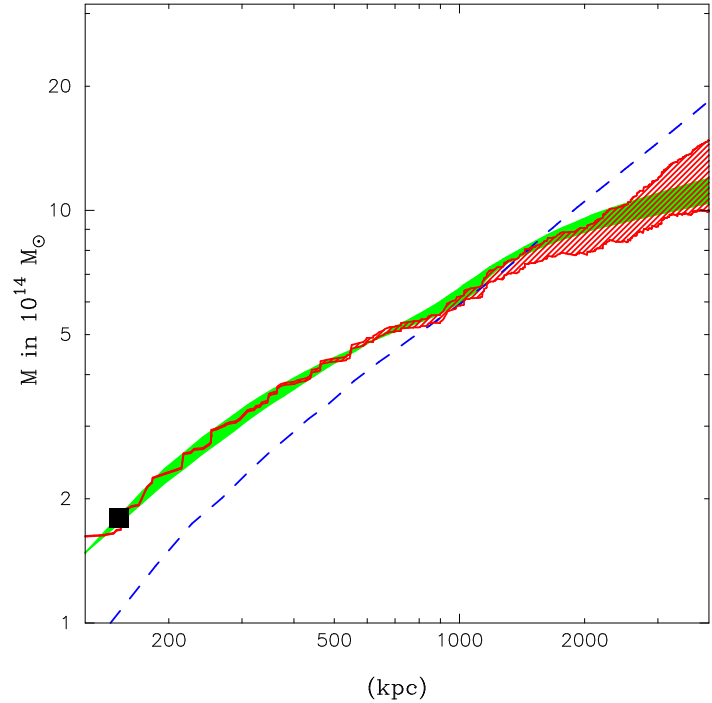}}
\figcaption{ Enclosed K-band light as a function of radius
(purple hatched region) scaled up by a constant M/L=40 
to roughly match the enclosed projected mass profiles.
The absolute mass contained within a 155 kpc
radius from the strong lensing model discussed in $\S$4.4 is shown
by the filled (red) square.
Uncertainties in the NFW fit are indicated by the light green region. 
The (blue) dashed line  shows the isothermal model that fits the
weak lensing data only.
}
\end{figure*}

\begin{figure*}
\centerline{\psfig{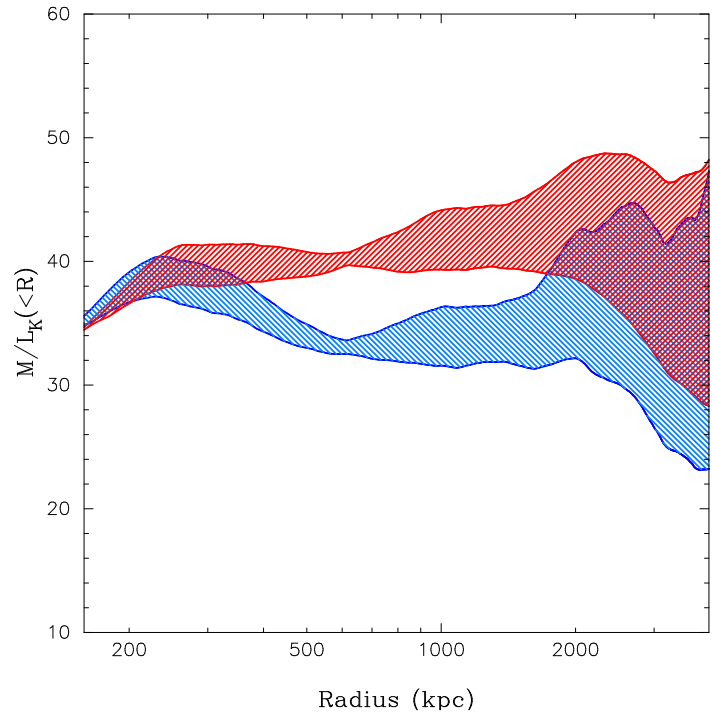}}
\figcaption{The degree to which radial variation in
the derived stellar mass/light ratio $M/L_K$ ratio (rest frame 
solar units) is permitted by the comparison of mass and light from 
the data shown in Figure~13. The (blue) lower hatched region corresponds to 
the $M/L$ derived for the enclosed field-subtracted $K$-band sample
and the (red) upper hatched region that for the color-selected sample.}
\end{figure*}

\section{Discussion}

In this paper we have used the exquisite image quality of {\it HST}
to derive the first radial profile of total mass
in a cluster to a projected radius of $\simeq$5 Mpc
combining strong and weak lensing constraints.
Our two main
results are that (i) the large scale distribution of total mass in
Cl0024+1654 over 0.1$<r<$5 Mpc is reasonably well fit 
by NFW-like profile (as opposed to the shallower isothermal form) 
and (ii) the mean mass/light ratio of the cluster
is constant over the same large dynamic range.

As far as the first result is concerned, the steep decline of the
mass density profile at large radii is an important constraint on
theories of cluster formation. Whereas at small radii the presence
of baryons and limited resolution of numerical simulations 
(Smith et al.\ 2001, Sand et al.\ 2002, Power et al.\ 2003) 
complicate comparisons between
data and models, these difficulties are negligible at large radii.
Our measured mass density profile declines as $r^{-n}$ with
$n>2.4$, clearly rejecting models that predict near-isothermal
slopes (see e.g. Ryden 1988). On large scales, our results are in
good agreement with the value $n=3$ obtained by NFW. 

Using a strategy very similar to that adopted
here utilizing a strong lensing constraint based on the
multiply-imaged arc, Bonnet et al.\ (1994) however inferred a projected 
mass density for a cored power law fit corresponding to $n\simeq$1.7-2.2. 
We argue that  the difference in the results comes
from the discrepant value in the shear estimate computed
by Bonnet et al.\ (1994), possibly resulting from over-correction of the
ground-based circularization of the faint galaxies, 
as discussed in section \S~4.

In considering how to make further progress in this area with
future cluster studies with {\it HST}, two issues are particularly
relevant. Firstly, our detection of substructure at radii 
$r\gtrsim 1.5$ Mpc was limited in this study by the sparse-sampling 
strategy adopted with WFPC2. While efficient in tracing the cluster 
profile to very large radius, it would be prudent in future studies to
undertake full sampling.
Secondly, there is an evident need to extend such studies to
further examples. Cl0024+1654 was found via optical searches which
rely on maximizing the contrast of galaxies against the
background. As often is the case, the most detailed analyses
reveal more complex structures than those foreseen in the earlier
data. For many years the high velocity dispersion was thought to
be inconsistent with the cluster's lensing power and the X-ray
luminosity. Czoske et al.\ (2001) resolved this discrepancy by
discovering a secondary peak (referred to as component B) in the
velocity distribution at $z\sim0.38$ as opposed to $z\sim0.395$ of
the main peak (A). The galaxies in component B are not spatially
concentrated but rather spread across the cluster field. This led
Czoske et al.\ (2002) to interpret the velocity distribution as
the result of a head-on collision between a moderate group and the
cluster proper.

Our lensing analysis likewise reveals an additional and
significantly massive secondary substructure $\sim 1$ Mpc NW of
the cluster center, spatially coincident with the secondary clump
of members of peak A in our spectroscopic catalog (Czoske et al.\ 2002,
Treu et al.\ 2003). Such substructure is likely to correlate with the high
fraction of blue galaxies in a cluster and consequently, as an
optically-selected system, Cl0024+1654 may not be completely
representative of massive ``relaxed" systems at this redshift,
in particular, unlike other relaxed clusters such as MS2137-23 
(Sand et al.\ 2002, Gavazzi et al.\ 2003) there is no central cD
in Cl0024+1654, but three giant elliptical galaxies.
Hence, Cl0024+1654 might not be the best cluster to measure the shape
of a Universal mass profile, as suggested by fit of ``merging-free"
clusters in numerical simulations.
Thus, understanding the mass profile in other more relaxed clusters 
would test whether our results can be generalized.

As far as the second result is concerned, the constant mass-to-light ratio
at large radii is remarkable both in terms of galaxy population and in
terms of relative spatial distributions of luminous and dark matter. 
In terms of galaxy population, this result adds support to the picture of
very gentle changes in the morphological mix and star formation histories
over the 0.1-5 Mpc range (Paper I).
In
terms of relative spatial distribution of luminous and dark matter,
their remarkable similarity and the implied constant $M/L$
ratio with radius, implies that dark matter and baryons are very
tightly coupled over a remarkable range of environmental
densities. Although at first surprising in term of simplistic
pictures of ``biased'' galaxy formation, the conclusion
strengthens one deduced completely independently in Paper I (Treu
et al.\ 2003). Clusters are primarily growing via the accretion of
groups, not individual galaxies. Accordingly, the peripheral
mass/light ratio represents not that of an individual galaxy
entering the cluster as an isolated ``test particle'' but rather
as part of a bound system whose overall mass/light ratio is fairly
high ($\simeq$30-40). The eradication of these smaller
substructures, which most likely occurs at and within the virial
radius (Paper I) should thus largely preserve the relative
distributions of mass and light on the scales which we can probe
via weak lensing.

\acknowledgments

We acknowledge useful discussions with Alexandre Refregier and
Tom Broadhurst. We thank
Chris Conselice and Kevin Bundy for assistance with the Palomar
WIRC infrared imaging and associated issues. 
We thank John Skilling of MaxEnt data consultants for providing the
{\sc BayeSys MCMC} software used in this work. We thank Paolo Tozzi 
for providing us with a reduced {\sl Chandra} image in advance of 
publication. We thank the anonymous referee for its useful comments.
This work was supported by NASA Grant. 
Jean-Paul Kneib acknowledges support
from CNRS and Caltech.  IRS acknowledges support from the Royal
Society and the Leverhulme Trust.
This paper is based on observations made with the
NASA/ESA Hubble Space Telescope, which is operated by the
Association of Universities for Research in Astronomy, Inc., under
NASA contract NAS~5-26555, and on observations 
obtained at the Canada-France-Hawaii
Telescope (CFHT) which is operated by the National Research
Council of Canada, the Institut National des Science de l'Univers
of the Centre National de la Recherche Scientifique of France,
and the University of Hawaii; as well as the Hale Telescope
at Mount Palomar using the WIRC camera.

\clearpage

\begin{deluxetable}{ccccccc}
\tablecaption{NFW fit parameters}
\tablewidth{0pt}
\tablehead{
\colhead{Clump} & \colhead{$x$} & \colhead{$y$}& \colhead{$M_{200}$}& \colhead{$c$}       &\colhead{$r_{200}$} &\colhead{$r_s$}}
\startdata
       & (arcsec)       & (arcsec)         &($10^{14}M_{\odot}$)&                     & (Mpc)      & (kpc) \\
1          & $0\pm1$        & $0\pm1$          & $6.1^{+1.2}_{1.1}$ & $22^{+9}_{-5}$      & $1.84\pm0.12$  & $83\pm 3$ \\
2          & $144^{+6}_{-9}$& $110^{+14}_{-8}$ & $2.6^{+1.4}_{-1.2}$& $4^{+2}_{-1}$ & $1.40^{+0.21}_{-0.24}$ & $340\pm 18$ \\
\enddata
\tablecomments{Parameters for the two
component NFW lens model described in the text, where we have applied
priors to the central clump to represent the strong lensing Einstein
radius constraint; the median sample values is used as a best
estimate. Uncertainties are 68\% probability symmetric intervals.
\label{tab:pars}}
\end{deluxetable}

\begin{deluxetable}{lccc}
\tablecaption{Posterior probability of Models Relative to NFW}
\tablewidth{0pt}
\tablehead{
\colhead{Type of Constraints} & \colhead{$Pr(CPL)/Pr(NFW)$} &
\colhead{$Pr(SIS)/Pr(NFW)$}& \colhead{$Pr(CIS)/Pr(NFW)$}}
\startdata
Weak Lensing Only & 2.5 & 1/15 & 1/400 \\
Strong+Weak Lensing & 2.5 & 1/9\,000 & 1/1\,800\,000 \\
\enddata

\tablecomments{Posterior probability of 2-component lens models
relative to the NFW model.  Given the weak lensing data alone (first
line), CPL and NFW are almost equally probable while the SIS and CIS
models are much less probable.  The second line corresponds to the
situation where strong lensing constraints are included as well. Now
SIS and CIS models are disfavored to a far greater degree. Note that
in both cases CPL is formally more probable than NFW, although not at
a significant level, given the additional free parameter.
\label{tab:probability}}
\end{deluxetable}

\end{document}